\documentclass[12pt,preprint]{aastex}
\usepackage{graphicx}

\begin{document}

\title{Chirality of Intermediate Filaments and Magnetic Helicity of Active Regions}
\author{Eun-Kyung Lim and Jongchul Chae}
\affil{Astronomy Program, Department of Physics and
Astronomy, Seoul National University, Seoul 151-742, Korea}
\email{uklim@astro.snu.ac.kr}

\begin{abstract}
Filaments which form either between or around active regions (ARs)
are called intermediate filaments. In spite of various theoretical
studies, the origin of the chirality of filaments is still
uncovered. We investigated how intermediate filaments are related to
their associated ARs, especially from the point of view of magnetic
helicity and the orientation of polarity inversion lines (PILs). The
chirality of filaments has been determined based on the orientations
of barbs observed in BBSO full-disk H$\alpha$ images taken during
the rising phase of solar cycle $23$. The sign of magnetic helicity
of ARs has been determined using \textsf{S}/inverse--\textsf{S}
shaped sigmoids from Yohkoh SXT images. As a result, we have found a
good correlation between the chirality of filaments and the magnetic
helicity sign of ARs. Among $45$ filaments, $42$ filaments have
shown the same sign as helicity sign of nearby ARs. It has been also
confirmed that the role of both the orientation and the relative
direction of PILs to ARs in determining the chirality of filaments
is not significant, against a theoretical prediction. These results
suggest that the chirality of intermediate filaments may originate
from magnetic helicity of their associated ARs.
\end{abstract}


\clearpage
\section{INTRODUCTION}
It is a well-known property that all filaments are located above the
polarity inversion lines (PILs) \citep{Bab55} and magnetic fields in
filaments are nearly parallel to PILs \citep{Ler78, Ler83}. The
direction of the parallel axial fields viewed from the positive
polarity side is the basis for defining the chirality of filaments
as either dextral or sinistral \citep{Mar98}. The alignment of
fibrils along the PIL in a filament channel observed even before the
visible filament forms \citep{Gai97}, indicates that the existence
of the highly sheared axial field components is a necessary
condition for the formation of filaments. Understanding how this
parallel axial field forms in a filament channel is the key to
understanding the formation of filament itself.

Solar differential rotation has been regarded as one of the most
probable mechanism, which introduces strong shear along the PILs
combined with other surface effects such as meridional flow and
diffusion \citep{van98, van00, Mac00, Marte01}. The common feature
of such models is the unexpected strong dependence of the chirality
on the initial orientation of PILs. According to \citet{van98}, in
the northern hemisphere, filaments which form along the east-west
oriented PILs have the sinistral chirality while filaments along the
north-south oriented PILs have the dextral chirality. As a result,
the similar numbers of dextral filaments and sinistral filaments
form in each hemisphere. This result contradicts the well-observed
hemispheric pattern of chirality reported by \citet{Mar94}. Then,
what caused this wrong prediction of chirality? Those models assumed
the initial condition of magnetic fields to be potential where there
is no initial twist before the formation of filaments. The surface
effects are the only source which generates the shear in PILs.
\citet{van00} and \citet{Mac00} also pointed out that this potential
field assumption cannot guarantee the sign of chirality which is
consistent with the observation. This recognition of the limitation
of the models based on the surface effects only raised the necessity
to find other origin of filament chirality that does not depend on
the initial geometric configuration such as the orientation of PILs.

The alternative source for the filament chirality suggested by
\citet{van00} and \citet{Mac00} is the injection of axial field
components along the PILs. Their simulation predicted the observed
chirality pattern correctly when there occured the axial flux
emergence during the evolution of magnetic fields. The estimated
amount of axial flux emergence is about $10^{19}$~Mx~day$^{-1}$.
However, do we see this amount of flux emergence on the quiet-sun
region prior to the formation of filaments? Unfortunately, there is
no observational evidence which supports this axial flux emergence
in the quiet-sun region. In spite of this lack of the observational
supports of axial flux emergence, the excellent agreement between
the predicted chirality from the simulation and the observed
chirality strongly indicates that the additional axial flux should
be provided somehow other than by the surface effects only.

Recent simulations have started considering the active region (AR)
magnetic helicity which is accumulated in the solar corona before
the filament formation as a probable origin of the filament
chirality. In an open volume such as corona, magnetic helicity could
be injected due to either the emergence of the helical flux through
the photosphere or the horizontal motion of the field lines on the
photosphere \citep{Ber84}. According to \citet{Jeo07}, helicity
injection in an AR occurs mainly at the early phase of its flux
emergence, which indicates that the majority of AR helicity is the
initial helicity and the helicity generated by surface effects is
the minority. The injected helicity is partly accumulated in the
corona and is partly lost by ejections via CMEs \citep{Dev00, Dem02,
Gre02, Lim07}. This AR helicity accumulated in the corona could
provide the initial twist to the filament channel which is essential
in order for axial fields to match the observed hemispheric pattern
and to reveal inverse polarity configuration also \citep{Mac03}. The
significant improvement of the research was carried out by
\citet{Mac03} who considered the initial helicity of ARs instead of
inserting axial flux into PILs during the evolution \citep{van00,
Mac00}. The importance of initial helicity in determining the
chirality of filaments was also suggested by \citet{Mac05} which
examined the variance of the skew angle according to both the
different sign of the initial helicity and the various tilt angles.
Their result shows the positive correlation between the chirality
and the sign of helicity within the range of tilt angle between
$20^{\circ}$ and $30^{\circ}$ in the northern hemisphere. On the
other hand, dextral skew was not produced for the large positive
tilt angle ($40^{\circ}$) even with the negative helicity sign, and
dextral skew was dominant within the range of tilt angle between
$-40^{\circ}$ and $10^{\circ}$ even with the positive helicity sign.
This result suggests that the surface effects still predominate over
the initial helicity under certain configuration.

Then which component is more fundamental and more important in the
formation of filament and chirality in the real solar space, surface
effects or initial helicity? According to our observational
experience, it is thought that the chirality of filaments is
determined predominantly by initial helicity contained in associated
ARs rather than the differential rotation. In order to verify the
significance of AR initial helicity in the filament formation and
its chirality determination in an observational way, we aim to
examine the one-to-one correspondence between the chirality of
quiescent filaments and the sign of the initial helicity of
associated ARs. In addition, we will also examine whether there is
any effect of the orientation of PILs on the chirality
determination.

As the first step of the observational study on quiescent filaments,
we limit the scope of the present work to focusing on the special
type of filaments, namely, intermediate. In that intermediate
filaments form external to single flux system, not within AR, we
regard intermediate filaments as a special type of quiescent
filaments. Unlike high latitude quiescent filaments that form
between two old flux systems, however, intermediate filaments form
adjacent to relatively young ARs. This property makes intermediate
filaments suitable objects to study the relationship between
filaments and their associated ARs.

\section{OBSERVATION \& METHOD}
We searched for intermediate filaments observed in the rising phase
of solar cycle $23$ from $1996$ to 2001, each of which formed
between a decayed polarity region and a relatively young AR.
Relatively short filaments near sunspots were selected as
intermediate filaments from BBSO H$\alpha$ full--disk images. Then
magnetic environments around filaments were examined using
line--of--sight full--disk magnetograms taken by SOHO/MDI, for the
purpose of confirming if they satisfy the above condition of
intermediate filaments.


The chirality of filaments has been determined based on the
orientation of barbs seen in the H$\alpha$ images that
right--bearing/left--bearing barbs always correspond to
dextral/sinistral \citep{Mar94}. Utilization of limb darkening
subtracted full--disk H$\alpha$ images provided by BBSO makes it
easy to check the orientation of barbs.
The uncertainty in
the determination of chirality of a filament depends on the number of barbs
as well as the seeing condition of H$\alpha$ images. If the seeing condition
is good enough and filaments have clear barbs more than 1,
the chirality can be unambiguously determined.
Most of intermediate filaments we examined  had three barbs, and that
chirality could be unambiguously determined unless the seeing condition is very poor.
Three filaments showed only one clear barb that was used to determine the chirality.
We have excluded from the sample a few filaments
that either ambiguously showed barbs of both type,
probably due to the projection effect, or whose structures were not clearly visible due to
the poor seeing condition of the images.


%

In order to determine the sign of magnetic helicity or twist of the
AR close to each intermediate filament, we used soft X--ray images
taken by SXT/Yohkoh. \citet{Can07} showed the one--to--one
correspondence between \textsf{S} or inverse--\textsf{S} sigmoids
within ARs and right--hand or left--hand twist of ARs respectively.
In case the shape of a sigmoid observed at the same day as a
filament is not clear, we checked SXT images one or two days before
or after the observation time assuming that the sign of helicity in
an AR maintains at least during a few days. Only 2 ARs contain
X--ray structures whose shape is neither S nor inverse--S.

\begin{figure}[tbp]
\begin{center}
    \includegraphics[width=0.32\textwidth]{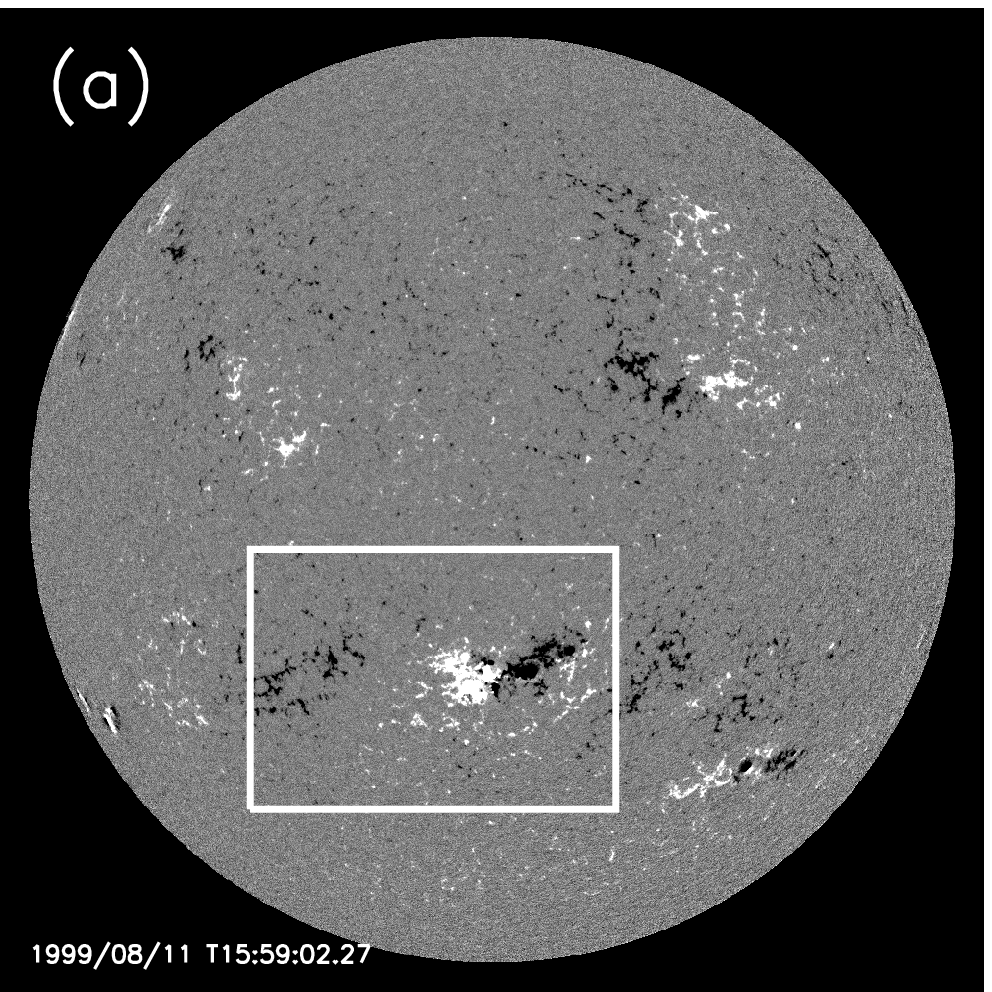}
    \includegraphics[width=0.32\textwidth]{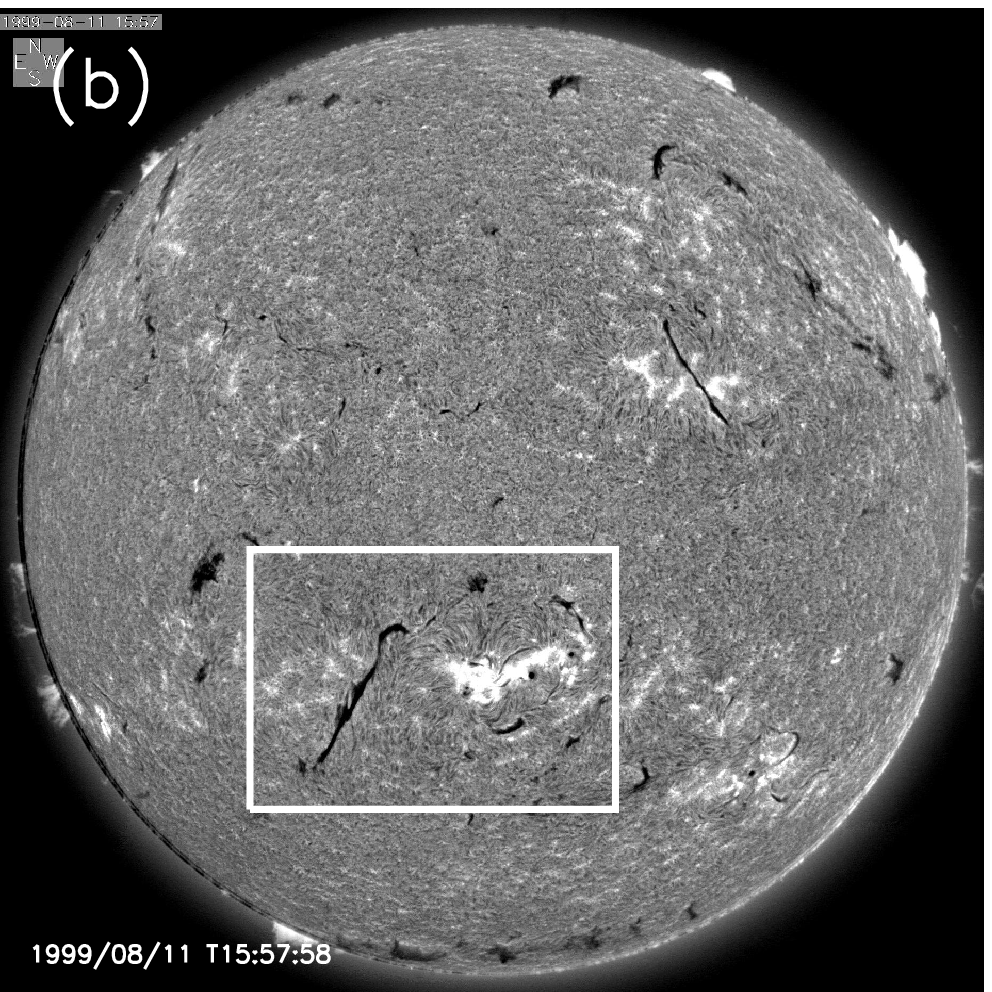}
    \includegraphics[width=0.32\textwidth]{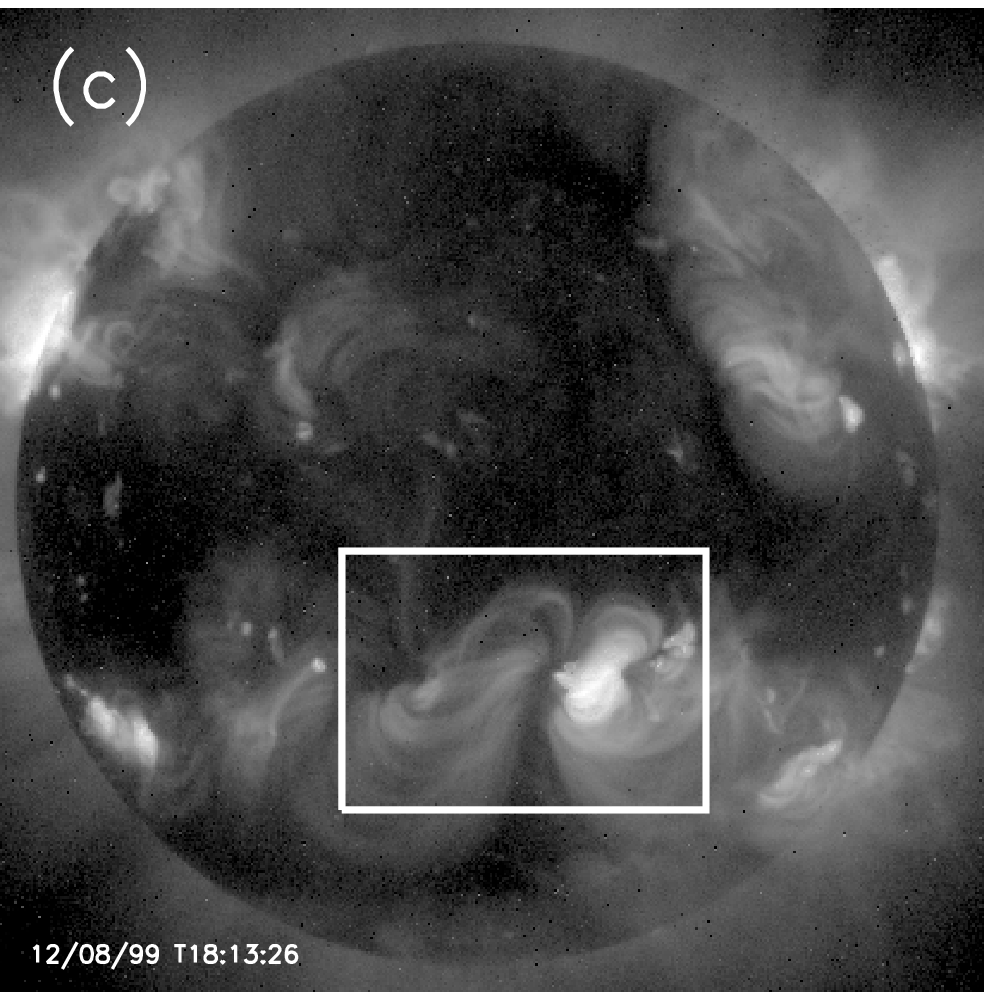}
\caption{A set of sample which contains the magnetogram, H$\alpha$
image and soft X-ray image. (a) MDI/SOHO magnetogram taken on 11
August 1999 (b) H$\alpha$ image taken on 11 August 1999 (c)
SXT/YOHKOH image taken on 12 August 1999}\label{fig1}
\end{center}
\end{figure}

Figure \ref{fig1} shows one filament in the sample. It was observed
on 11 August 1999; SOHO/MDI magnetogram, BBSO/H$\alpha$ image and
Yohkoh/SXT image. It is clear that the filament in the southern
hemisphere is an intermediate filament since it formed between the
negative polarity of decaying flux region in the left side and the
positive polarity of the younger AR in the right side
(Figure~\ref{fig1}(a)). The left-bearing barbs shown in
Figure~\ref{fig1}(b) indicate that the filament is sinistral. In
this case of sinistral filament in the southern hemisphere, the
orientation of the PIL is nearly north-south direction and the AR is
in the western side of the PIL. Therefore, we sort this filament
into group 'N-S' in terms of the PIL orientation, and group 'W' in
terms of the relative direction of the AR. The compact and bright
coronal structure observed in the SXT image (Figure~\ref{fig1}(c))
which is in the same location as the AR shows \textsf{S} shaped
sigmoid indicating that the twist of the AR is right-hand sense, or
the sign of the helicity in this AR is positive.

\section{RESULT}

\subsection{Examples}

\begin{figure}[tbp]
\begin{center}
\plottwo{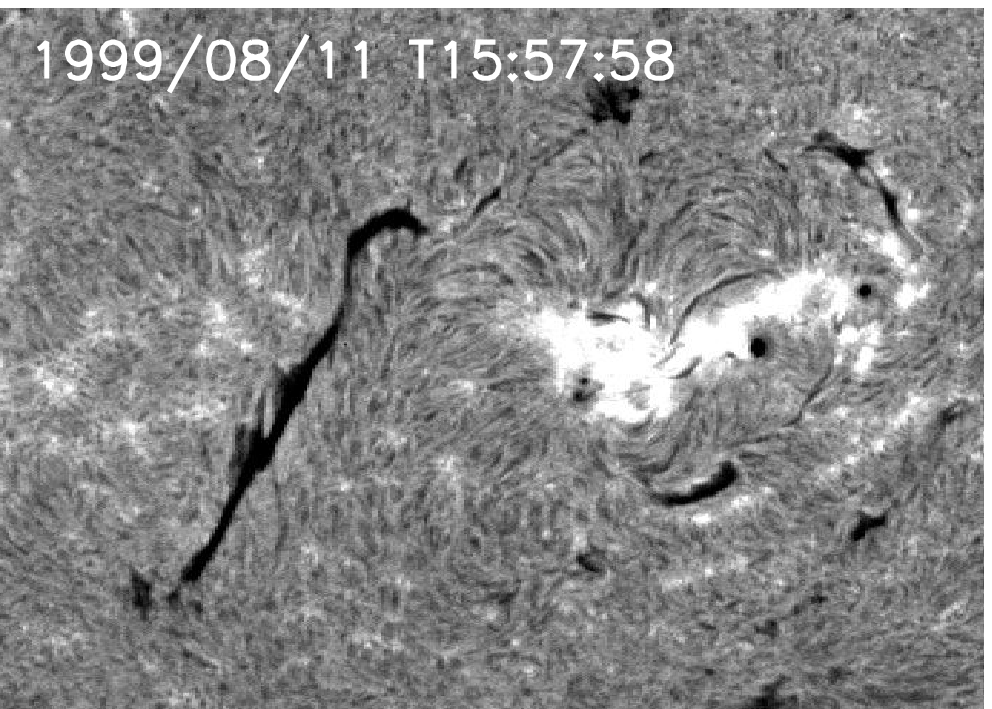}{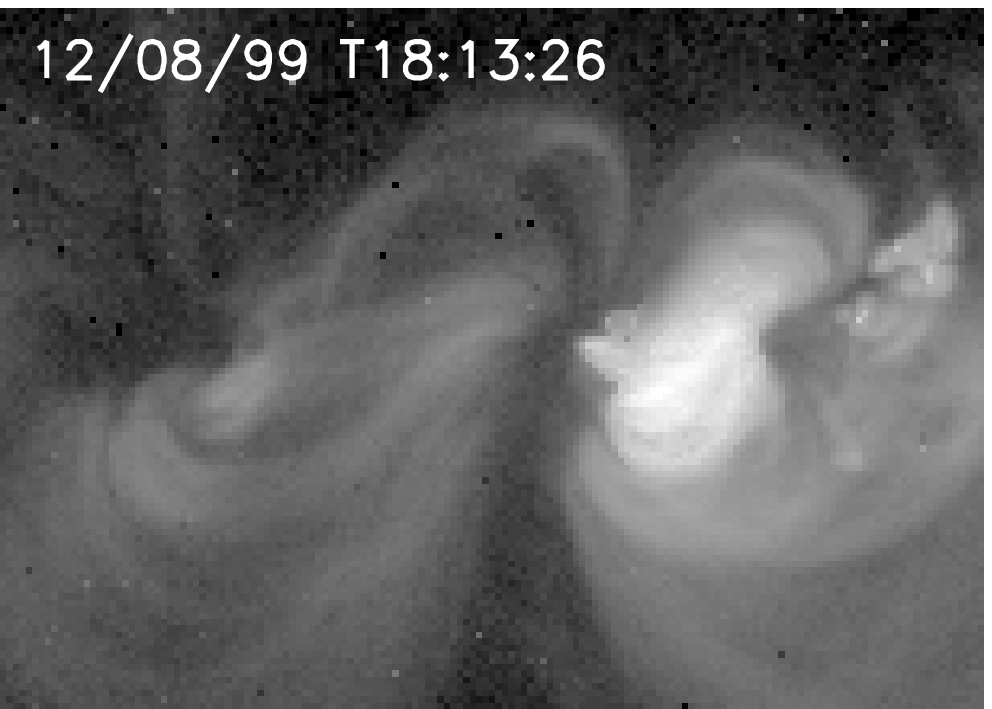}
\caption{A sinistral filament in the southern hemisphere. The
\textsf{S} sigmoid above the adjacent AR indicates the positive AR
helicity.} \label{fig2}
\end{center}
\end{figure}

\begin{figure}[tbp]
\begin{center}
\plottwo{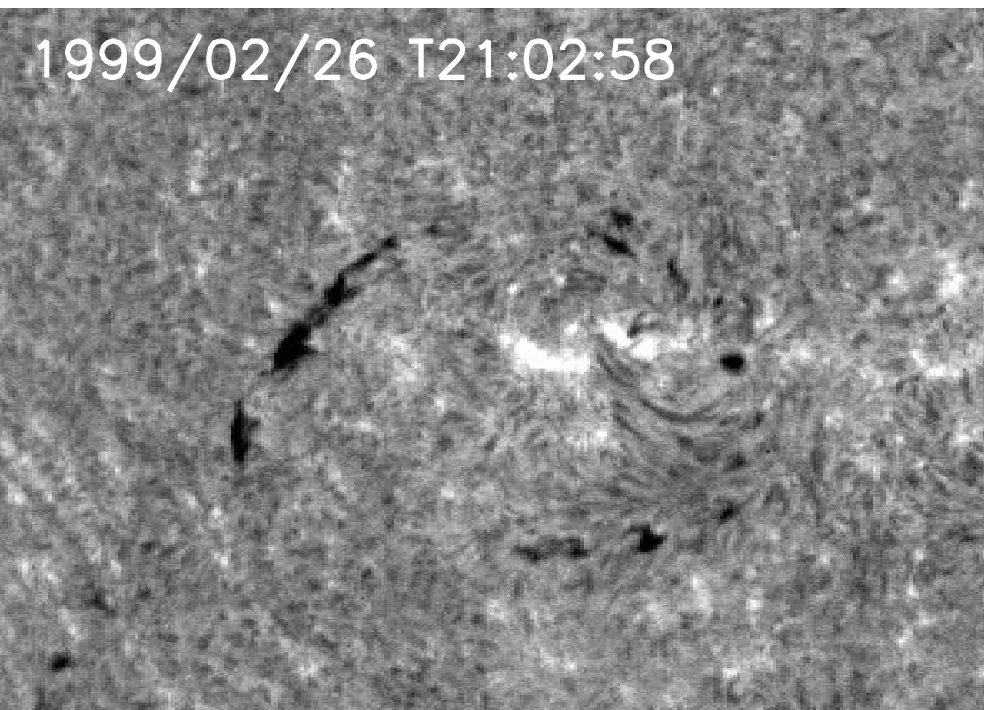}{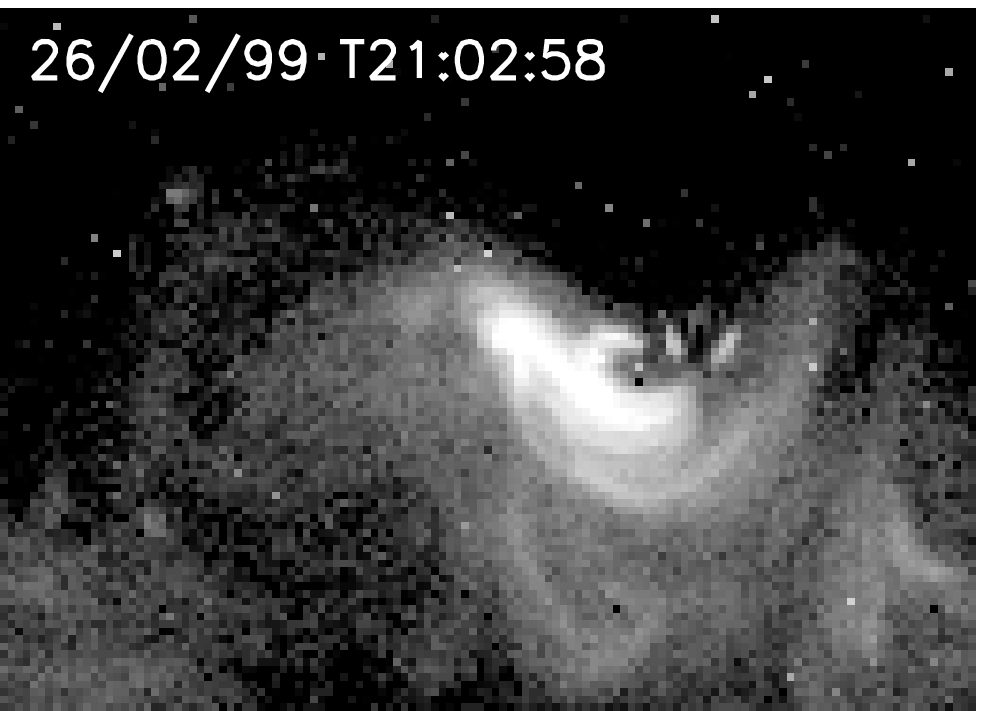}
\caption{A dextral filament in the southern hemisphere. The
inverse-\textsf{S} sigmoid above the adjacent AR indicates the
negative AR helicity.} \label{fig3}
\end{center}
\end{figure}

Figure~\ref{fig2} and Figure~\ref{fig3} are examples of sinistral
and dextral filaments observed on 11 August 1999 and on 26 February
1999, respectively. Figure~\ref{fig2} is a partially enlarged image
of Figure~\ref{fig1}. Both filaments in Figure~\ref{fig1} and
Figure~\ref{fig3} are located in the southern hemisphere and in the
eastern side of each AR. However, as seen in the SXT images, the
signs of their associated ARs are the opposite. In case of the AR
observed on 11 August 1999, Figure~\ref{fig2}, the shape of the
bright sigmoid is \textsf{S}. Two bundles of loops in the north and
south part of the AR form \textsf{S} shaped sigmoid which means this
AR contains the positive helicity.  The helicity of this active
region is the same as the statistically determined hemispheric
pattern of active region helicity.
 On the other hand, the AR
observed on 26 February 1999 shows a definite inverse-\textsf{S}
shaped sigmoid which indicates the presence of the negative helicity
in the AR. This is opposite to the hemispheric pattern mentioned above.
The point is that  even though these two ARs are in the same hemisphere,
and in the same relative direction to the filament, they have  opposite signs of
helicity. More important is that  the associated filaments have opposite chirality
signs, too; the filament adjacent to the positive-helicity AR is sinistral and
the one adjacent to the negative-helicity AR is dextral.
This suggests that the relationship between AR helicity and filament chirality
is not a coincidence resulting from the hemispheric patterns of AR helicity and
filament chirality.

One of the well known features of filaments is the existence of the
right-skewed coronal arcades above the sinistral filaments and
left-skewed coronal arcades above the dextral filaments
\citep{Mar97, Mar98}. This feature is also notable in
Figure~\ref{fig2} and Figure~\ref{fig3}. The coronal arcade above
the filaments are anchored in magnetic pole of AR next to the
filaments. This close linkage between AR magnetic fields and magnetic
environment of the filaments indicates that the intermediate filaments may have
formed under the magnetic environment that is strongly affected by the adjacent ARs.
As a consequence, it may not be surprising that  the chirality of an intermediate
filament is related to the helicity of the associated AR.

\subsection{Statistics}

Among the filaments observed during the rising phase of cycle 23, we
have selected $45$ filaments the chirality of which were reliably determined
and whose associated ARs showed clear sigmoid coronal
structures. These filaments were grouped by their chirality,
helicity sign of their nearby ARs, the orientations of PILs and the
relative directions of adjacent ARs to old flux regions.

%

 Table~\ref{tbl1} shows that the relationship between the chirality of filaments  and the sign of AR helicity.  It indicates that 38 among 45 filaments (84\% level) follow the hemispheric pattern, and 39 among 45 ARs (87\%) follow the hemispheric pattern. The table shows that there is a very
strong correlation between the filament chirality and the sign of AR
helicity in each hemisphere. About  42 among 45 intermediate filaments ($93\%$) have the same sign of chirality as that of their associated AR helicity. This
strong correlation between the chirality and the AR helicity sign
supports the idea that there is one-to-one correspondence between
the chirality of filaments and the sign of  AR helicity.
Especially in the southern hemisphere,
all sinistral filaments are associated with the positive AR helicity,
and all dextral filaments with negative AR helicity. This strong
correlation may not be due to the statistically  established hemispheric
patterns of filament chirality  and active region helicity,
for dextral filaments in the southern hemisphere do not follow such a pattern.

On the other hand, in the northern hemisphere, there are
three filaments, one dextral and two sinistral, which do not
follow the strong correlation as seen in the southern hemisphere.
Do these filaments represent clear counter examples indicating any exception
to the inferred relationship between filament chirality and AR helicity sign?
They may be not. As for the dextral filament,  the determination of the sign of AR helicity was somewhat ambiguous, because the sigmoid of AR was not so well developed.
In the other two sinistral filaments,  the associated ARs were complex,  not being well isolated from other ARs. These complex magnetic environments surrounding the filaments may have introduced the ambiguity in identifying the associated active region.
 Therefore, we think that these three cases represent some uncertainty in our approach  rather than clear counter examples.
If we can find any clear counter example which does not follow the correlation, that would be quite interesting one, but so far we have found none.



\begin{table}[tbp]
\begin{center}
\caption{The relationship between the sign of chirality and the sign
of AR helicity. Numbers in bold represent the number of filaments
which have the positive correlation between chirality and AR
helicity. \label{tbl1}}\vspace{1ex}
\vspace{1ex}
\begin{tabular}{c|cc}
\multicolumn{3}{c}{(a) Northern hemisphere}\\
\tableline\tableline
\rule{0pt}{3ex}AR helicity & \multicolumn{2}{|c}{Filament chirality}\\
\rule{0pt}{3ex}            & (sinistral) & (dextral)\\
\tableline
\rule{0pt}{3ex} $+$ & $\textbf{3}$ & $1$ \\
\rule{0pt}{3ex} $-$ & $2$ & $\textbf{19}$ \\
\tableline
\rule{0pt}{3ex}Total & $5$ & $20$ \\
\tableline
\end{tabular}\,\,\,\,\,
\vspace{3ex}
\begin{tabular}{c|cc}
\multicolumn{3}{c}{(b) Southern hemisphere}\\
\tableline\tableline
\rule{0pt}{3ex}AR helicity & \multicolumn{2}{|c}{Filament chirality}\\
\rule{0pt}{3ex}            & (sinistral) & (dextral)\\
\tableline
\rule{0pt}{3ex} $+$ & $\textbf{18}$ & $0$ \\
\rule{0pt}{3ex} $-$ & $0$ & $\textbf{2}$ \\
\tableline
\rule{0pt}{3ex}Total & $18$ & $2$ \\
\tableline
\end{tabular}
\end{center}
\end{table}

Table~\ref{tbl2} suggests that the PIL orientation may not be
important in determining the chirality of filaments, unlike some
theoretical studies based on the surface effects \citep{van98,
van00}. According to these studies, it is expected that the 'N-S'
orientation of the PIL corresponds to dextral (sinistral) chirality
in the northern (southern) hemisphere. Table~\ref{tbl2} indicates
that $31$ filaments are consistent with this prediction, but
$6$ filaments are not. We think that this number ($6$ out of
$31$) may not be statistically negligible. On the other hand,
the studies predict that the 'E-W' orientation of the PIL
corresponds to sinistral (dextral) chirality in the northern
(southern) hemisphere. However, Table~\ref{tbl2} shows that only one
out of seven filaments is consistent with this prediction, and the
other six filaments are not. Therefore, our results strongly suggest
that the orientation of the PIL may not be the most important factor
in the determination of filament chirality, and hence differential
rotation may not be crucial to the chirality of filaments, even
though it might play some role.

\begin{table}[tb]
\begin{center}
\caption{The relationship between the orientation of PILs and the
chirality of filaments. NH (SH) indicates the northern (southern)
hemisphere. The numbers of filaments consistent with the theoretical
 prediction (see text) are indicated in bold.
\label{tbl2}} \vspace{2ex}
\begin{tabular}{c|cc}
\tableline\tableline
\rule{0pt}{3ex}PIL orientation & \multicolumn{2}{|c}{Filament chirality}\\
\rule{0pt}{3ex} & dextral(NH)/sinistral(SH) & sinistral(NH)/dextral(SH) \\
\tableline
\rule{0pt}{3ex}N---S & $\textbf{32}$ & $6$ \\
\rule{0pt}{3ex}E---W & $6$ & $\textbf{1}$ \\
\tableline
\rule{0pt}{3ex}Total & $39$ & $7$ \\
\tableline
\end{tabular}\\
\end{center}
\end{table}

Next, we have compared our result with the model of \citet{Mac98}
which predicted that the position of the younger AR relative to the
older, decaying flux system, is one factor affecting the filament
chirality. Table~\ref{tbl3} shows the result we have obtained to
test the model prediction. Note that for easy comparison, only the
filaments with the 'N-S' PIL were used to produce the table. In the
northern hemisphere, only $40\%$ of filaments are consistent
with the model and in the southern hemisphere, $47\%$ are,
clearly indicating that the relative position of the younger AR is
not an important factor unlike the prediction.

\begin{table}[tb]
\begin{center}
\caption{The relationship between the chirality and the relative
position of AR to the older flux systems in each hemisphere. The
numbers of filaments consistent with the prediction of \citet{Mac98}
are indicated in bold. \label{tbl3}} \vspace{1ex}
%
\vspace{1ex}
\begin{tabular}{c|cc}
\multicolumn{3}{c}{(a) Northern hemisphere}\\
\tableline\tableline
\rule{0pt}{3ex}AR position & \multicolumn{2}{|c}{Filament chirality}\\
\rule{0pt}{3ex}              & (sinistral) & (dextral)\\
\tableline
\rule{0pt}{3ex}eastern side & $\textbf{2}$ & $10$ \\
\rule{0pt}{3ex}western side & $2$ & $\textbf{6}$ \\
\tableline
\rule{0pt}{3ex}Total & $4$ & $16$ \\
\tableline
\end{tabular}\,\,\,\,\,
%
\vspace{3ex}
\begin{tabular}{c|cc}
\multicolumn{3}{c}{(b) Southern hemisphere}\\
\tableline \tableline
\rule{0pt}{3ex}AR position & \multicolumn{2}{|c}{Filament chirality}\\
\rule{0pt}{3ex}              & (sinistral) & (dextral)\\
\tableline
\rule{0pt}{3ex}eastern side & $7$ & $\textbf{0}$ \\
\rule{0pt}{3ex}western side & $\textbf{8}$ & $2$ \\
\tableline
\rule{0pt}{3ex}Total & $15$ & $2$ \\
\tableline
\end{tabular}
\end{center}
\end{table}

\section{DISCUSSION}
Our observation of intermediate filaments using BBSO H$\alpha$
full-disk images during the rising phase of solar cycle $23$ bears
out three main results: 1) There is a strong correlation between the
sign of filament chirality and the sign of AR helicity. 2) The
orientation of the PIL is not crucial to determining the filament
chirality. 3) The position of the associated ARs relative to the old
flux systems is not correlated to the sign of filament chirality.
The second result implies that the major origin of filament
chirality can not be differential rotation, the effect of which
helicity injection is sensitive to the orientation of the PIL. In
this regard, our result supports the studies done by Mackay and his
collaborators \citep{van00, Mac00, Mac03}. Our result on the
relative position of ARs, however, does not support \citet{Mac98}
which attempted to the formation and chirality of an intermediate
filament.

These results also raise a challenge to all of previous efforts to
explain the chirality of filaments by assuming geometric factors
such as orientation, location, specific configuration and so on.
These assumptions  were often stated implicitly in the form of
schematic drawings.  For example, \citet{Marte01} successfully
reproduced the hemispheric pattern of the filament chirality using
the head-to-tail linkage model of two initially unconnected bipoles.
The success of their model is attributed to the configuration where
the bipole at the lower latitude is slightly displaced toward the
west with respect to the one at the higher latitude. There is no
explanation why this configuration should happen. It might be due to
the effect of differential rotation.   If this is  the case,the
origin of filament chirality in their model is the differential
rotation,  which  our results do not support.

Another example of a schematic model is found in the recent work of
\citet{Wan07} that reported the observation of the formation of
intermediate filaments. They used the magnetic reconnection between
potential-like global magnetic fields of the AR and the  bipoles
located on the outer polarity inversion line, and then successfully
produced dextral filaments as observed (see Fig. 10 in the paper).
Note that the success of their model is due to their choice
of the PIL in the upper left. If they had chosen the PIL on the
upper right, the filament would have turned out to be sinistral.
Therefore this model suggests that it is possible for both dextral filaments and sinistral ones to be associated with the same active region if it has a potential-like field configuration. This prediction is not consistent with our results indicating the strong correlation between the chirality of filaments and the sign of AR helicity. The discrepancy
might be attributed to the fact that the ARs we examined were far from potential-like configurations as employed by \citet{Wan07}. It would then be interesting to examine whether ARs of weakly twisted or potential field configuration can have intermediate filaments of both signs of chirality or not.

The most important result we have obtained, of course, is the strong
correlation between the sign of chirality of intermediate filaments
and that of helicity of their associated ARs. This result may shed
light on the physical nature of the {\em initial helicity} that has
been often assumed to explain for the correct sign of filament
chirality \citep{van00, Mac00, Mac03, Mac05}. Since a quiescent
filament is most likely to be formed at the interface of two flux
systems when they converge \citep{Tan87}, the initial helicity of
the filament should come from either of two flux systems. If one of
these flux systems is younger than the other, which is the case for
intermediate filaments, the helicity of filaments may be more
contributed by the younger one as we shall argue below. According to
this scenario, the sign of chirality of intermediate filaments
should be equal to the sign of AR magnetic helicity, being
consistent with our result.

In the case of an intermediate filament, the helicity of the younger
system  is likely to dominate over the older flux system  since the
helicity content of a flux system  may decrease with its age. The
intensive injection of helicity in ARs occurs mainly in the flux
emerging phase, and the total amount of helicity injected during
this period is about a few $10^{43}\,\textrm{Mx}^2$ \citep{Jeo07}.
Once the notable flux emergence is over so that the injection of
helicity becomes mere, the rest of the lifetime may be regarded as a
helicity-losing period. CMEs, for instance, are a well-known
mechanism through which helicity is expelled from the corona of ARs.
Each CME carries out helicity with an amount of
$\sim10^{42}\,\textrm{Mx}^2$ \citep{Dev00}. Supposing
 that over $5-10$ times of CMEs occur
from an AR during its lifetime \citep{Dev00, Dem02}, a very old flux
system may contain little amount of remnant helicity. On the other
hand, a young flux system may have experienced less CMEs, so that it
may keep enough amount of helicity that can contribute to the
initial helicity for the formation of filaments. As a consequence,
the contribution of the younger flux system would be dominant.

Supposing that our above speculation on the intermediate filaments is correct and  it can be extended to quiescent filaments in a similar way,  we present a prediction on the formation and chirality of quiescent filaments which form between two old flux systems. If the signs of remnant helicity of two old flux systems
are the same, then either or both of two systems would supply
helicity into the PIL, leading to the formation of filaments with
the same sign of chirality. On the other hand, if the two systems
contain helicity of opposite sign, which system will contribute more
would not be determined straightforward. Then we could think of two
possibilities. If the two systems have remnant helicity of
comparable amount, but of different sign, the net helicity at the
interface would  be close to zero, and filaments would not be formed
there. If, on the other hand, there is unbalance between the amount
of helicity, the interface would contain significant amount of
helicity and filaments would be formed. The sign of chirality of
these quiescent filaments would follow that of the dominant
helicity. We plan to investigate the relationship between the
chirality of quiescent filaments and AR remnant helicity to verify
this prediction.

\acknowledgements We greatly appreciate the referee's constructive comments.  This work was supported by the Korea Research
Foundation Grant funded by the Korean Government
(KRF-2005-070-C00059), and the Seoul Fellowship.


\begin{thebibliography}{}
%
\bibitem[Babcock \& Babcock\,(1955)]{Bab55}
Babcock,~H.~W. and Babcock,~H.~D. 1955, ApJ, 543, 447
%
\bibitem[Berger \& Field\,(1984)]{Ber84}
Berger,~M.~A., and Field,~G.~B., 1984, J.~Fluid~Mech., 147, 133
%
\bibitem[Canfield et al.\,(2007)]{Can07}
Canfield,~R.~C., Kazachenko,~M.~D., Acton,~L.~W., Mackay,~D.~H.,
Son,~J., and Freeman,~T.~L. 2007, ApJ,671,L81
%
\bibitem[D\'{e}moulin et al.\,(2002)]{Dem02}
D\'{e}moulin,~P., Mandrini,~C.~H., van~Driel-Gesztelyi,~L.,
Thompson,~B.~J., Plunkett,~S., K\H{o}v\'{a}ri,~Zs., Aulanier,~G.,
and Young,~A. 2002, A\&A, 382, 650
%
%
\bibitem[DeVore\,(2000)]{Dev00}
DeVore,~C.~R. 2000, ApJ, 539, 944
%
\bibitem[Gaizauskas et al.\,(1997)]{Gai97}
Gaizauskas,~V., Zirker,~J.B., Sweetland,~C., and Kovacs,~A. 1997,
ApJ, 479, 448
%
\bibitem[Green et al.\,(2002)]{Gre02}
Green,~L.~M., L\'{o}pez Fuentes,~M.~C., Mandrini,~C.~H.,
D\'{e}moulin,~P., van~Driel-Gesztelyi,~L., and Culhane,~J.~L. 2002,
Sol.~Phys., 208, 43
%
\bibitem[Jeong \& Chae\,(2007)]{Jeo07}
Jeong,~H. and Chae.~J. 2007, ApJ, 671, 1022
%
\bibitem[Lim et al.\,(2007)]{Lim07}
Lim,~E.~-K., Jeong,~H., and Chae,~J. 2007, ApJ, 656, 1167
%
\bibitem[Leroy\,(1978)]{Ler78}
Leroy,~J.~L. 1978, A\&A, 64, 247
%
\bibitem[Leroy et al.\,(1983)]{Ler83}
Leroy,~J.~L., Bommier,~V., and Sahal-Brechot,~S. 1983, Sol.~Phys.
83, 135
%
%
\bibitem[Mackay et al.\,(1998)]{Mac98}
Mackay,~D.~H., Priest,~E.~R., Gaizauskas,~V., and
van~Ballegooijen,~A.~A. 1998, Sol.~Phys., 180, 299
%
\bibitem[Mackay et al.\,(2000)]{Mac00}
Mackay,~D.~H., Gaizauskas,~V., and van~Ballegooijen,~V. 2000, ApJ,
544, 1122
%
\bibitem[Mackay\,\&\,Gaizauskas\,(2003)]{Mac03}
Mackay,~D.~H., and Gaizauskas,~V. 2003, Sol.~Phys., 216, 121
%
\bibitem[Mackay\,\&\,van~Ballegooijen\,(2005)]{Mac05}
Mackay,~D.~H. and van~Ballegooijen,~A.~A. 2005, ApJ, 621, L77
%
%
\bibitem[Martens \& Zwaan\,(2001)]{Marte01}
Marten,~P.~C., and Zwaan,~C. 2001, ApJ, 558, 872
%
\bibitem[Martin, Bilimoria, and Tracadas\,(1994)]{Mar94}
Martin,~S.~F., Bilimoria,~R., and Tracadas,~P.~W. 1994, in Solar
Surface Magnetism, ed. R.~J.~Rutten and C.~J.~Schrijver (NATO ASI
Ser. C, 433; Dordrecht: Kluwer), 303
%
\bibitem[Martin\,\&\,McAllister\,(1997)]{Mar97}
Martin,~S.~F., and McAllister,~A.~H. 1997, in Coronal Mass Ejections,
ed. N. Crooker, J.Joselyn, \& J. Feynman (Geophys. Monogr. 99;
Washington, DC: AGU), 127
%
\bibitem[Martin\,(1998)]{Mar98}
Martin,~S.~F. 1998, ASP conf., 150, 419M
%
%
%
\bibitem[Tang\,(1987)]{Tan87}
Tang,~F. 1987, Sol.~Phys., 107, 233
%
\bibitem[van~Ballegooijen et al.\,(1998)]{van98}
van~Ballegooijen,~A.~A., Cartledge,~N.~P., and Priest,~E.~R. 1998,
ApJ, 501, 866
%
\bibitem[van~Ballegooijen et al.\,(2000)]{van00}
van~Ballegooijen,~A.~A., Priest,~E.~R., and Mackay,~D.~H. 2000, ApJ,
539, 983
%
\bibitem[Wang \& Muglach\,(2007)]{Wan07}
Wang,~Y.~-M., and Muglach,~K. 2007, ApJ, 666, 1284
%
%
\end{thebibliography}
\end{document}